\documentclass[submission,copyright,creativecommons]{eptcs}
\providecommand{\event}{GandALF 2020}

\usepackage{etex}

\usepackage[utf8]{inputenc}
\usepackage[T1]{fontenc}
\usepackage{anyfontsize}

\usepackage{microtype}
\usepackage[british]{babel}
\usepackage{soul}

\usepackage[framemethod=tikz]{mdframed}
\usepackage{wrapfig}

\usepackage{amssymb}
\usepackage{amsmath}
\usepackage{amsfonts}
\usepackage{stmaryrd}

\usepackage{calrsfs}
\DeclareMathAlphabet{\pazocal}{OMS}{zplm}{m}{n}

\usepackage{amsthm}
\newtheorem{example}{Example}

\usepackage{tabu,booktabs}
\usepackage{multicol,multirow}
\usepackage[inline,shortlabels]{enumitem}

\usepackage{tikz}
\usetikzlibrary{calc, patterns, shapes, shapes.misc, intersections, positioning, arrows.meta}
\usetikzlibrary{decorations.pathreplacing, backgrounds}
\usetikzlibrary{hobby}
\usetikzlibrary{automata, arrows, calc, shapes, shapes.misc, shapes.multipart, decorations.pathreplacing, decorations.markings, decorations.pathmorphing, fit, patterns, positioning}

\usepackage[noend]{algorithm2e}
\AlgoDontDisplayBlockMarkers
\DontPrintSemicolon
\LinesNumbered
\SetAlgoCaptionLayout{normal}
\SetAlgoCaptionSeparator{}
\SetAlCapSkip{.5em}
\makeatletter
\renewcommand{\algocf@makecaption}[2]{%
	\addtolength{\hsize}{1.5\algomargin}%
	\parbox[t]{\hsize}{\algocf@captiontext{#1:}{#2}}%
	\addtolength{\hsize}{-1.5\algomargin}%
}
\makeatother
\SetFuncArgSty{textup}
\SetArgSty{textup}

\SetCommentSty{mycommfont}
\SetKwComment{tcp}{}{}
\SetKwProg{Def}{def}{:}{}
\SetKwProg{Thread}{thread}{:}{}
\SetKwIF{If}{ElseIf}{Else}{if}{:}{elif}{else:}{}
\SetKwFor{For}{for}{:}{}
\SetKwFor{ForAll}{forall}{:}{}
\SetKwFor{While}{while}{:}{}
\SetKwIF{Catch}{}{Try}{catch}{:}{}{try:}{}
\SetKwFor{Loop}{loop:}{}{}
\SetKwFor{DoPar}{do in parallel:}{}{}
\SetKw{Break}{break}
\SetKw{Raise}{raise}
\SetKw{False}{false}
\SetKw{True}{true}
\SetKw{Continue}{continue}
\SetKw{KwTo}{in}
\SetKwFunction{DFI}{dfi}
\SetKwFunction{Winner}{winner}
\SetKwFunction{Onestep}{onestep}

\renewcommand\implies{\Rightarrow}

\newcommand{\pr}{\textsf{pr}}

\newcommand{\invalpha}{{\overline{\alpha}}}

\newcommand{\sqdiamond}{\tikz [x=1.2ex,y=1.2ex,line width=.08ex] \draw (0,.5) -- (.5,1) -- (1,.5) -- (.5,0) -- (0,.5) -- cycle;}
\newcommand{\sqsq}{\tikz [x=0.95ex,y=1ex,line width=.1ex] \draw (0,0) -- (1,0) -- (1,1) -- (0,1) -- (0,0) -- cycle;}

\newcommand{\Even}{{\raisebox{0.13ex}{\scalebox{0.95}{\sqdiamond}}}}
\newcommand{\Odd}{{\raisebox{0.15ex}{\scalebox{0.9}{$\sqsq$}}}}

\mdfdefinestyle{todo}{tikzsetting={draw=blue,fill=blue!10,line width=3pt},leftmargin=0pt,
	rightmargin=0pt,innertopmargin=6pt,innerbottommargin=6pt,innerleftmargin=6pt,innerrightmargin=6pt}
\newenvironment{TodoBoxedInternal}[1][]
{\begin{mdframed}[style=todo]\parindent0pt}{\end{mdframed}}

\iftrue
\newcommand{\Todo}[1]{%
	\begin{TodoBoxedInternal}
		\textbf{TODO} \quad #1
	\end{TodoBoxedInternal}%
}
\else
\newcommand{\Todo}[1]{}
\fi

\definecolor{Gray}{HTML}{8c8c8c}
\newcommand{\stdevtext}[1]{}

\title{Symbolic Parity Game Solvers that Yield Winning Strategies}
\def\titlerunning{Symbolic Parity Game Solvers that Yield Winning Strategies}

\author{
	Oebele Lijzenga
	\institute{Formal Methods and Tools \\ University of Twente, Enschede}
	\email{o.r.lijzenga@student.utwente.nl}
	\and
	Tom van Dijk
	\institute{Formal Methods and Tools \\ University of Twente, Enschede}
	\email{t.vandijk@utwente.nl}
}
\def\authorrunning{Oebele Lijzenga \& Tom van Dijk}

\begin{document}

\maketitle
\begin{abstract}
	Parity games play an important role for LTL synthesis as evidenced by recent breakthroughs
	on LTL synthesis, which rely in part on parity game solving.
	Yet state space explosion remains a major issue if we want to scale to larger systems or specifications.
	In order to combat this problem, we need to investigate symbolic methods such as BDDs, which have been successful in the past to tackle exponentially large systems.
	It is therefore essential to have symbolic parity game solving algorithms, operating using BDDs, that are fast and that can produce the winning strategies used to synthesize the controller in LTL synthesis.

	Current symbolic parity game solving algorithms do not yield winning strategies.
	We now propose two symbolic algorithms that yield winning strategies, based on two recently proposed fixpoint algorithms.
	We implement the algorithms and empirically evaluate them using benchmarks obtained from SYNTCOMP 2020.
	Our conclusion is that the algorithms are competitive with or faster than an earlier symbolic implementation of Zielonka's recursive algorithm, while also providing the winning strategies.
\end{abstract}

\section{Introduction}
\label{section:introduction}
Parity games are turn-based games played by the players \textit{Even} and \textit{Odd} on a finite directional graph. All vertices are labelled by an integer priority. A play in a parity game is an infinite sequence of vertices consistent with the edge relation, where the owner of the current vertex (\textit{Even} or \textit{Odd}) determines the next vertex. The play eventually results in an infinitely repeating sequence of vertices and their priorities. If the highest priority in this sequence is even, then player \textit{Even} wins, otherwise player \textit{Odd} wins. Solving a parity game can either be to determine the winning area of the graph for a player, or determining a winning strategy for both players, or both. A strategy is winning for some player if it contains one move for each vertex controlled by, and in the winning area of that player, and all moves consistent with that strategy always cause that player to win.

It is widely believed that a polynomial time solution exists for determining the winner of a parity game, as they lie in the intersection of UP and co-UP~\cite{DBLP:journals/ipl/Jurdzinski98}, which is contained in the intersection of NP and co-NP.
In the current paper, however, we are not interested in theoretical complexity but in practical performance.
Parity games are closely related to many problems in formal verification and synthesis that can be reduced to the problem of solving parity games, as parity games capture the expressive power of nested least and greatest fixpoint operators.
Parity games are especially relevant for LTL synthesis.
%
%
In recent years, the fastest practical solver for LTL synthesis has been \textsc{Strix}~\cite{DBLP:conf/cav/MeyerSL18}, winning SYNTCOMP~\cite{DBLP:journals/corr/abs-1711-11439} editions 2018~and~2019. \textsc{Strix} converts an LTL formula to a deterministic parity automaton and then splits the automaton into a parity game.
The solution of the parity game determines if the LTL formula is realizable and the winning strategy is used to construct a controller that is guaranteed to implement the LTL specification.

There are various classes of parity game solving algorithms.
Broadly speaking, we identify four classes of algorithms.
First, the algorithms that use repeated attractor computation to partition a game into regions, which are subsequently refined until the game is solved.
A particular algorithm in this category is Zielonka's recursive algorithm~\cite{DBLP:journals/tcs/Zielonka98}, where the partitioning is done recursively and subgames are fully solved before refining the top region.
Recent work in this category includes novel algorithms like priority promotion~\cite{DBLP:journals/fmsd/BenerecettiDM18} and tangle learning~\cite{DBLP:conf/cav/Dijk18}, but also a quasi-polynomial variation of the recursive algorithm~\cite{DBLP:conf/mfcs/Parys19}.
Second, the class of progress measures algorithms~\cite{DBLP:conf/stacs/Jurdzinski00}, where each vertex in the game has a value which increases monotonically, based on the values of the direct successors of the vertex.
These values are typically tuples of integers, and in recent work various authors have proposed sets of values that are quasi-polynomially bounded in size yet sufficient to solve parity games~\cite{DBLP:conf/stoc/CaludeJKL017,DBLP:conf/soda/CzerwinskiDFJLP19}.
Third, many different strategy improvement algorithms~\cite{DBLP:conf/cav/VogeJ00} have been studied, where one player iteratively improves its strategies by playing against the best response of the opponent.
Fourth, several algorithms that do not fit in the above categories are formulated as a nested fixpoint in the modal $\mu$-calculus.
We call these simply the fixpoint iteration algorithms.
These are the \textsc{APT} algorithm~\cite{stasio_bdd_parity}, a fixpoint iteration algorithm we refer to as \textsc{BFL}~\cite{DBLP:journals/corr/BruseFL14}, as well as the recent \textsc{DFI}~\cite{DBLP:journals/corr/abs-1909-07659} and \textsc{FPJ}~\cite{DBLP:conf/vmcai/LapauwBD20} algorithms.


Often formal verification and also synthesis suffers from the \emph{state space explosion} problem.
Indeed, \cite{willemse} reports that parity game solvers can require over 8 GB memory for large specifications.
One method to alleviate this problem is to use binary decision diagrams~\cite{DBLP:journals/iandc/BurchCMDH92}.
%
In the past, symbolic parity game solvers using binary decision diagrams have been explored~\cite{DBLP:conf/mochart/BakeraEKR08,kant_parity_bdd,willemse,stasio_bdd_parity}.
Symbolic algorithms replace explicit data structures with implicit, symbolic representations such as binary decision diagrams, relying on optimised BDD implementations such as CUDD~\cite{DBLP:journals/sttt/Somenzi01}
and Sylvan~\cite{DBLP:journals/sttt/DijkP17}.
Using binary decision diagrams can result in a massive difference in memory usage~\cite{willemse}.

For applications such as LTL synthesis, but also for model checking, obtaining the winning strategies is essential.
Controller synthesis requires the winning strategy. 
For model checking, the winning strategy is used to derive counterexamples.
In the current literature, several symbolic implementations of parity game solving algorithms have been proposed.
In \cite{kant_parity_bdd}, Kant et al. implement Zielonka's recursive algorithm symbolically.
Their focus lies on generating the games and little attention is given to strategy derivation.
Their online implementation in LTSmin~\cite{DBLP:conf/tacas/KantLMPBD15} appears to be capable of generating the winning strategy, but this is not reported in the paper and the solver is embedded within the LTSmin model checker.
A symbolic implementation of the quasi-polynomial ``ordered progress measures'' algorithm~\cite{DBLP:journals/sttt/FearnleyJKSSW19} has been proposed~\cite{DBLP:conf/lpar/ChatterjeeDHS18} which requires only quasi-polynomially many symbolic operations.
In \cite{stasio_bdd_parity}, Di Stasio et al. implement Zielonka's recursive algorithm, the fixpoint algorithm APT \cite{apt}, and a symbolic small progress measures~\cite{DBLP:conf/stacs/Jurdzinski00} algorithm.
Winning strategies are not derived.
Finally, Sanchez et al.~\cite{willemse} implement Zielonka's recursive algorithm, two fixpoint iteration algorithms, and priority promotion~\cite{DBLP:journals/fmsd/BenerecettiDM18} symbolically, but without strategy derivation.

If we want to use symbolic methods in the entire LTL synthesis toolchain, from specification to controller, then we require fast symbolic parity game algorithms that produce the winning strategies.




In recent work, Van Dijk and Rubbens~\cite{DBLP:journals/corr/abs-1909-07659} and Lapauw et al.~\cite{DBLP:conf/vmcai/LapauwBD20} propose fixpoint algorithms that yield strategies in a straightforward way.
Previous fixpoint algorithms either do not produce a winning strategy~\cite{apt} or require a secondary, complicated algorithm~\cite{DBLP:journals/corr/BruseFL14}.
Fixpoint iteration algorithms have the advantage that the data structures that they maintain are very simple.
For various algorithms like strategy improvement and small progress measures, relatively complex and refined labelings are used,
whereas fixpoint iteration algorithms maintain very little information per vertex: membership in the distraction set is sufficient, and in addition the algorithms we study also store the strategy (a subset of the edge relation) and either the so-called justification (a subset of the edge relation), or membership in one of the $d$ so-called frozen sets, where $d$ is the highest priority in the parity game.

The contributions of this paper are as follows.
We \textbf{propose} and \textbf{implement} two novel symbolic parity game algorithms based on \cite{DBLP:journals/corr/abs-1909-07659} and \cite{DBLP:conf/vmcai/LapauwBD20}.
We \textbf{evaluate} these algorithms empirically using benchmarks from the 2020 edition of SYNTCOMP~\cite{DBLP:journals/corr/abs-1711-11439}.
We find that the algorithms are competitive compared to state-of-the-art algorithms, while also producing winning strategies.

\section{Preliminaries}
\label{section:preliminaries}
\subsection{Parity games}
We define a parity game PG as a tuple $(V, V_\Diamond, V_\Box, E, \pr)$, where $V$ is a set of vertices partitioned into $V_\Diamond$ and $V_\Box$; vertices in $V_\Diamond$ are controlled by player Even and vertices in $V_\Box$ are controlled by player Odd.
The mapping $E \subseteq V \times V$ describes which moves can be made from each vertex in $V$; each vertex must have at least one successor.
The function $\pr$: $V \to \{0,1,...,d\}$ assigns a priority to each vertex, where $d$ is the highest priority in the game.


We introduce some additional notation.
We write $E(u)$ for all successors of $u$, and $u \to v$ if $v \in E(u)$.
We write player $\alpha \in \{\Diamond, \Box\}$ when referring to one of the players and player $\invalpha$ to refer to the other player.
We use $V_0$ for vertices with an even priority and $V_1$ for vertices with an odd priority.
We extend this notation for any set $X\subseteq V$ such that $X_0 := X\cap V_0$, $X_\alpha := X\cap V_\alpha$, etc.
%
%
%
Finally, we use the following notations from $\mu$-calculus~\cite{mu_calculus} to denote predecessors with \emph{some} or \emph{all} successors in the set $X$:
\[
\begin{tabu} to \linewidth {rrl}
\Diamond X & := & \{ v \in V \mid \exists u \colon v \to u \land u \in X\} \\
\Box X & := & \{ v \in V \mid \forall u \colon v \to u \implies u \in X\} \\
\end{tabu}
\]
With $\Diamond X$ we mean all vertices with a successor in $X$.
This is also called the \emph{preimage} of $X$. 
With $\Box X$ we mean the vertices where \emph{all} successors are in $X$, i.e., vertices with no edges to vertices outside $X$.

A \emph{play} $\pi=v_0 v_1 \dots$ is an infinite sequence of vertices consistent with $E$, i.e.,
$v_i \rightarrow v_{i+1}$ for all successive vertices.
We denote with $\inf(\pi)$ the vertices that occur infinitely often in $\pi$.
Player Even wins a play $\pi$ if the highest priority in $\inf(\pi)$ is even; player Odd if the highest priority in $\inf(\pi)$ is odd.
%

A (positional) \emph{strategy} $\sigma\colon V\to V$ assigns to each vertex in its domain a single successor in $E$, i.e., $\sigma\subseteq E$.
We refer to a strategy of a player $\alpha$ to restrict the domain of $\sigma$ to $V_\alpha$. 
In the remainder, all strategies $\sigma$ are of a player $\alpha$.
We write $\text{Plays}(v)$ for the set of plays starting at vertex $v$ and
$\text{Plays}(v, \sigma)$ for all plays from $v$ consistent with $\sigma$.

A basic result for parity games is that they are memoryless determined~\cite{DBLP:conf/focs/EmersonJ91}, i.e., each vertex is either winning for player Even or for player Odd, and both players have a positional strategy for their winning vertices.
Player $\alpha$ wins a vertex $v$ if there exists a strategy $\sigma$ of player $\alpha$ such that every $\pi\in\text{Plays}(v,\sigma)$ is winning for player $\alpha$.

\begin{example}
Figure~\ref{fig:parity_game} is an example of a simple parity game, consisting of 9 vertices and 15 edges.
The diamond-shaped vertices are in $V_\Diamond$ and the square-shaped vertices are in $V_\Box$.
Player Odd wins all vertices of the parity game with the strategy $\{ \textbf{b}\to\textbf{f}, \textbf{d}\to \textbf{e}\}$.
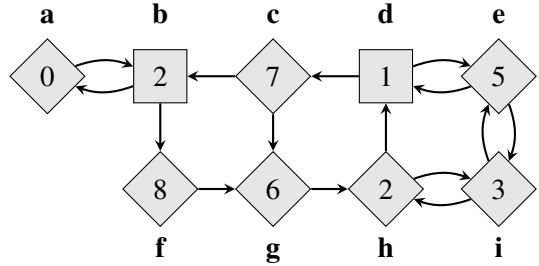
\begin{figure}[btp]
    \centering
	\begin{tikzpicture}
	\tikzset{every edge/.append style={>=stealth,->,solid,thick,draw,text height=0.5ex,text depth=0.2ex}}
	\tikzset{every node/.append style={inner sep=0,minimum size=10mm,draw,fill=black!10,font={\upshape}}}
	\tikzset{my label/.style args={#1:#2}{
			append after command={
				($(\tikzlastnode.center)$) coordinate [label={[label distance=3mm,black]#1:\textbf{\strut #2}}]
	}}}
	\tikzstyle{even}=[diamond]
	\tikzstyle{odd}=[regular polygon,regular polygon sides=4]
	
	\draw (-1.5,1.5)  node[even, my label={above:a}] (a) {0};
	\draw (0,1.5)     node[odd,  my label={above:b}] (b) {2};
	\draw (1.5,1.5)   node[even, my label={above:c}] (c) {7};
	\draw (3,1.5)     node[odd,  my label={above:d}] (d) {1};
	\draw (4.5,1.5)   node[even, my label={above:e}] (e) {5};
	\draw (0,0)       node[even, my label={below:f}] (f) {8};
	\draw (1.5,0)     node[even, my label={below:g}] (g) {6};
	\draw (3,0)       node[even, my label={below:h}] (h) {2};
	\draw (4.5,0)     node[even, my label={below:i}] (i) {3};
	
	\draw (c) edge (b);
	\draw (f) edge (g);
	\draw (b) edge (f);
	\draw (b) edge [bend left=20] (a);
	\draw (a) edge [bend left=20] (b);
	\draw (g) edge (h);
	\draw (c) edge (g);
	\draw (d) edge [bend left=20] (e);
	\draw (e) edge [bend left=20] (d);
	\draw (d) edge (c);
	\draw (h) edge [bend left=20] (i);
	\draw (i) edge [bend left=20] (h);
	\draw (h) edge (d);
	\draw (e) edge [bend left=20] (i);
	\draw (i) edge [bend left=20] (e);
	\end{tikzpicture}
    \caption{A simple parity game, consisting of 9 vertices and 15 edges.}
    \label{fig:parity_game}
\end{figure}
\end{example}

\subsection{Fixpoint iteration algorithms for parity games}

Many different types of algorithms have been proposed to solve parity games.
In this paper, we focus on fixpoint iteration algorithms.
Fixpoint iteration algorithms iteratively refine an estimation of which vertices are won by which player.
A typical initial estimation is that player Even wins all even-priority vertices and player Odd wins all odd-priority vertices.
This estimation is then iteratively refined by considering the direct successors of each vertex.
By updating the estimation in a strict order, starting with the lowest-priority vertices, and resetting the estimation of lower-priority vertices whenever a higher-priority vertex is updated, the solution of parity games can be computed~\cite{DBLP:journals/corr/abs-1909-07659}.

We consider two fixpoint iteration algorithms proposed in recent literature.
The \texttt{DFI} algorithm proposed by Van Dijk and Rubbens~\cite{DBLP:journals/corr/abs-1909-07659} is a fixpoint iteration algorithm based on the concept of \textit{distractions}.
The freezing extension of \texttt{DFI} enables straightforward strategy computation.
When updating vertices of some priority, all lower-priority vertices that are won by the winner are frozen, i.e., they will no longer be reevaluated.
Only vertices that are not frozen are reevaluated.
The effect is that freezing preserves the winning strategy.
The \texttt{FPJ} algorithm proposed by Lapauw et al.~\cite{DBLP:conf/vmcai/LapauwBD20} modifies the standard fixpoint iteration by maintaining a \emph{justification graph}, which essentially records which vertices are currently ``justified'' and the strategy that witnesses the justification.
Justified vertices are not reevaluated, but each time a vertex is updated by the algorithm, the justification graph is pruned by removing vertices that are no longer justified.
It is straightforward to extract the winning strategy from the justification graph.

A distraction is some even-priority vertex $v$ that can be won by player Odd if player Even always tries to reach $v$, and vice versa.
We say that a distraction is \emph{fatal} if player Odd wins vertex $v$ regardless of the strategy of player Even, and that a distraction is \emph{devious} if player Even wins vertex $v$ using a strategy that at least sometimes avoids vertex $v$.
They are called distractions because (many) parity game solvers will initially assume that these vertices are safe for player $\alpha$ to play to, but after some steps, the parity game solver will find that these vertices are actually unsafe for player $\alpha$ and have to recompute all vertices that relied on the unsafe vertex.
For example in Figure~\ref{fig:parity_game}, vertex \textbf{c} is a devious distraction.
If player Odd plays from \textbf{d} to \textbf{c}, then player Even wins. By avoiding vertex \textbf{c}, player Odd wins the game.

Fixpoint algorithms offer a distinct advantage compared to other parity game algorithms, as they maintain uncomplicated evaluations of vertices.
Earlier work, e.g.~\cite{willemse,stasio_bdd_parity}, also considers fixpoint algorithms in their symbolic implementations.
As demonstrated in~\cite{stasio_bdd_parity}, the symbolic implementation of the small progress measures algorithm often results in timeouts.

\subsection{Binary decision diagrams}

\begin{figure}[tbp]
\centering
	\tikzstyle{var}=[ellipse,draw=black!70,inner sep=0.5mm,solid,thick,minimum size=6mm]%
	\tikzstyle{leaf}=[rectangle,draw=black!70,solid,thick,minimum size=6mm]%
	\tikzstyle{p}=[->,>=stealth',solid,thick]%
	\tikzstyle{n}=[->,>=stealth',dashed,thick]%
	\begin{tabu} to 0.8\linewidth {X[c]|X[c]|X[c]|X[c]}
		\toprule
		$x$ & $x_1\wedge x_2$ & $x_1\vee x_2$ & $x_1 \oplus x_2$ \\
		\midrule
		\begin{tikzpicture}
		\node[var] (x) {$x$}
		child[p] {node[leaf] (T) {$1$} }
		child[n] {node[leaf] (F) {$0$} };
		\end{tikzpicture}
		&
		\begin{tikzpicture}
		\node[var] (X) {$x_1$}
		child[p] {node[var] (Y) {$x_2$}
			child[p] {node[leaf] (T) {$1$} }
			child[n] {node[leaf] (F) {$0$} }
		};
		\draw[n, bend left=30] (X) to (F);
		\end{tikzpicture}
		&
		\begin{tikzpicture}
		\node[var] (X) {$x_1$}
		child[n] {node[var] (Y) {$x_2$}
			child[p] {node[leaf] (T) {$1$} }
			child[n] {node[leaf] (F) {$0$} }
		};
		\draw[p, bend right=30] (X) to (T);
		\end{tikzpicture}
		&
		\begin{tikzpicture}
		\node[var] (X) {$x_1$}
		child[p] {node[var] (Y1) {$x_2$}
			child[n] {node[leaf] (T)  {$1$} }}
		child[n] {node[var] (Y2) {$x_2$}
			child[n] {node[leaf] (F)  {$0$} }};
		\draw[p] (Y2) to (T);
		\draw[p] (Y1) to (F);
		\end{tikzpicture}
		\\
		\bottomrule
	\end{tabu}
\caption{Four simple binary decision diagrams.}
\label{fig:bdds}
\end{figure}
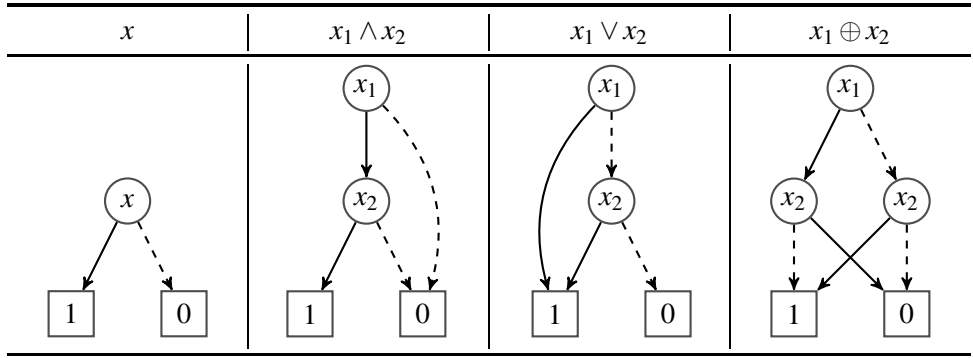

Binary decision diagrams \cite{DBLP:journals/csur/Bryant92,DBLP:books/daglib/0029044,DBLP:journals/sttt/DrechslerS01} (BDDs) are a well known data structure for representing and manipulating Boolean functions.
Figure \ref{fig:bdds} shows four examples of BDDs representing different Boolean functions.
A binary decision diagram is a rooted directed acyclic graph, with internal nodes representing decisions over Boolean variables.
Each internal node has two successors, one via the so-called ``true'' edge (depicted as a solid arrow) and one via the ``false'' edge (depicted as a dashed arrow).
Given some valuation to Boolean variables, we either follow the ``true'' edge if the variable has the value true, or the ``false'' edge.
If we arrive at the leaf node ``1'', then the represented Boolean function evaluates to true for the given valuation.
Otherwise, the represented Boolean function evaluates to false.
It is well known that BDDs are a canonical representation of Boolean functions if they are ordered, i.e., Boolean variables are encountered according to a fixed variable ordering, and reduced, i.e., redundant decision nodes with two identical successors are removed~\cite{DBLP:journals/csur/Bryant92}.

Given some vector of $N$ Boolean variables $\vec{x}$ and a decision diagram representing some function $f$, binary decision diagrams can very concisely represent sets $S\subseteq 2^N$, where $S = \{ x \mid f(x) = 1\}$ is the set of valuations to $\vec{x}$ for which the represented Boolean function $f$ evaluates to true.
One can obtain a BDD for any set by means of a binary encoding of the elements of the set and the application of the Boole-Shannon decomposition $S = x\cdot S_x+\bar{x}\cdot S_{\bar{x}}$, where $S_x$ is the subset of $S$ where $x$ equals 1, and $S_{\bar{x}}$ is the subset of $S$ where $x$ equals 0~\cite{boole1854investigation,shannon1938symbolic}.
It is well known that BDDs can be incredibly efficient if a suitable variable ordering is found and the represented set is encoded in a way that results in small decision diagrams.
Further study of variable ordering and encoding lie outside of the scope of the current paper and we refer to~\cite{DBLP:journals/sttt/DrechslerS01} for a comprehensive overview of binary decision diagrams.

\section{Symbolic fixpoint algorithms that yield winning strategies}
\label{section:bdds}
We now present our approach to converting the explicit parity game algorithms to their symbolic counterparts.
In order to solve a parity game symbolically, the parity game itself has to be represented symbolically using BDD.
Furthermore, data structures and procedures from \texttt{DFI} and \texttt{FPJ} have to be modified in order to fit the symbolic paradigms.
Due to space constraints, we do not present the explicit algorithms \texttt{DFI} and \texttt{FPI} here and instead refer to~\cite{DBLP:journals/corr/abs-1909-07659} and~\cite{DBLP:conf/vmcai/LapauwBD20}.


\subsection{Encoding of the parity game}

To use decision diagrams as the data structure to store parity games, we replace explicit data structures by their symbolic counterparts.
We use a naive binary encoding.
Sets are modelled as a functions over a vector $\Vec{x}$.
Our input parity games have vertices numbered from $0\dots n$ and we simply encode vertices by encoding these numbers.
For example we use $4$ Boolean variables to represent a parity game with $9$ vertices.
Notice that in a full LTL synthesis toolchain, we would retain more information about each vertex and could do better than a naive binary encoding. This is a topic of future work.

We have a BDD for the set $V$ of all vertices in the parity game.
Furthermore, we have BDDs for $V_\Diamond$, $V_\Box$ and a BDD for $V^p$ for every priority $p$.
We thus choose to represent the priority function using one BDD for every priority.
An alternative could be to use a multi-terminal BDD (or algebraic decision diagram) for the priority function, but practical parity games have only few priorities anyway.




Finally, to represent the successor relation $E$, we introduce additional Boolean variables $\vec{x}'$. We use the original variables $\vec{x}$ for the source vertex and the new variables $\vec{x}'$ for the target vertex.
We order all $\vec{x}$ before all $\vec{x}'$ in the variable ordering.
%
%

\subsection{Symbolic alternatives for explicit procedures}

When manipulating binary decision diagrams, the internal nodes are never modified directly.
Instead, BDD operations compute new decision diagrams, reusing existing nodes as much as possible, thus ensuring that BDDs are canonical representations.
As a consequence, computing something like a union $S := S \lor T$ results in a new BDD, that may or may not share nodes with the original representations of $S$ and $T$.
When we perform more complicated operations, such as $S := S \land (A \lor B \lor C)$, the order of operations may have a major influence on the size of intermediate BDDs.

%

The original algorithms as presented in e.g.~\cite{DBLP:journals/corr/abs-1909-07659} use {for}-loops over individual vertices.
In order to have efficient BDD implementations, this must be avoided whenever possible.
The power of symbolic data structures lies in representing and manipulating sets of vertices rather than individual vertices.

When algorithms use data structures such as arrays that assign some value to each vertex, we need a symbolic representation of such arrays.
If only few values are assigned, then we can represent these arrays using one BDD for each possible value.
Alternatively we could encode the assigned values or use multi-terminal BDDs.
For example we use sets $V^p$ for every priority in the game, since there are only few priorities anyway.
If the parity game had unique priorities for every vertex, such a representation would be very inefficient.
The representation may also influence operations that might be wasteful when operating on a set encoding all values, and efficient when operating on sets representing the value that we are interested in.
Hence, the choice of representation is not trivial, and may have significant impact on the performance of algorithms.

Two particular procedures are computing the sets $\Diamond X$ and $\Box X$, i.e., predecessors of $X$:
\[
\begin{tabu} to \linewidth {rrl}
\Diamond X & := & \{ v \in V \mid \exists u \colon v \to u \land u \in X\} \\
\Box X & := & \{ v \in V \mid \forall u \colon v \to u \implies u \in X\} \\
\end{tabu}
\]

Computing the preimage $\Diamond X$ 
is a well known operation.
We first substitute the Boolean variables of the target set $X$, which is represented using Boolean variables $\vec{x}$, to use Boolean variables $\vec{x}'$ instead.
Then we intersect the successor relation $E$ with this set $X'$.
By performing existential quantification of the $\vec{x}'$ variables on the intersection, we obtain all vertices that have a successor in the original set $X$.

Computing $\Box X$ is slightly different.
We obtain the set $X'$ as for $\Diamond X$.
Now we compute $v \to u \implies u \in X \equiv \neg (v\to u) \lor (u \in X)$ simply by adding $X'$ to the complement of $E$: $\neg E \lor X'$: the result is the set of edges that either do not exist or whose successor is in $X$; any edge not in this set is an existing edge to a vertex outside $X$.
Hence, if we now perform universal quantification of the $\vec{x}'$ variables on this intersection, we obtain all vertices without successors outside $X$, i.e., $\Box X$.

\subsection{Symbolic \texttt{DFI}}

\begin{algorithm}[tbp]
    \Def{\texttt{\textup{dfi}}(PG)}{
		$Z \gets \emptyset$ \tcp*{initialize distractions set}
		$F_0 \gets \emptyset , \dotsc , F_d \gets \emptyset$ \tcp*{initialize each freezing set}
		$S \gets \emptyset$ \tcp*{initialize the strategy to be empty}
		$p \gets 0$ \tcp*{we start updating with the lowest priority}
		\While(\tcp*[f]{run until all fixpoints are stable}){$p\leq d$}{
			$\alpha \gets p \bmod 2$ \tcp*{get the current player}
			$X \gets ((((V^p \land \neg Z) \land\neg F_0)\land\neg F_1)\dotsc)\land \neg F_d$ \tcp*{compute unfrozen vertices of priority $p$}
			\lIf(\tcp*[f]{compute now-distractions (winning for 1)}){$\alpha=0$}{
			    $Z' \gets X \land\neg \texttt{onestep}_0(X, Z)$ 
			}
			\lElse(\tcp*[f]{compute now-distractions (winning for 0)}){
			    $Z' \gets \texttt{onestep}_0(X, Z)$ 
			}

			$Z \gets Z \vee Z'$ \tcp*{add new distactions to $Z$}
			$S \gets (S \wedge \neg X)$ \tcp*{reset strategy for unfrozen vertices}
			$S \gets S \lor (X \wedge V_\Even \wedge E \wedge \texttt{subst}(\texttt{even}(Z), \vec{x}\to\vec{x}'))$ \tcp*{add strategy for Even to \texttt{even}}
			$S \gets S \lor (X \wedge V_\Odd \wedge E \wedge \texttt{subst}(\texttt{odd}(Z), \vec{x}\to\vec{x}'))$ \tcp*{add strategy for Odd to \texttt{odd}}
			
			\If(\tcp*[f]{did we get new distractions?}){$Z' \neq \emptyset$}{			    
			    $X \gets ((V^0 \lor V^1)\dotsc)\lor V^{p-1}$ \tcp*{get all vertices $<p$}
			    $X \gets (((X\land\neg F_0)\land\neg F_1)\dotsc)\land \neg F_d$ \tcp*{... that are not frozen}
			    \lIf(\tcp*[f]{... and are won by even}){$\alpha=0$}{
			    	$W \gets X \land \texttt{even}(Z)$ 
		    	}
   				\lElse(\tcp*[f]{... and are won by odd}){
   					$W \gets X \land \texttt{odd}(Z)$
   				}
			    $F_p \gets F_p \lor (X \land \neg W)$ \tcp*{freeze lower vertices won by $\invalpha$: $X\setminus W$}
			    $Z \gets Z \land \neg W$ \tcp*{reset lower vertices won by $\alpha$: $W$}
			    $p \gets 0$ \tcp*{continue with lowest priority}
			}
			\Else(\tcp*[f]{no new distractions?}){
			    $F_p \gets \emptyset$ \tcp*{thaw vertices}
			    $p \gets p + 1$ \tcp*{continue with next priority}
			}
		}
	    $W_\Even \gets \texttt{even}(Z)$ \tcp*{winning region for Even}
		$W_\Odd \gets \texttt{odd}(Z)$ \tcp*{winning region for Odd}
		$S_\Even \gets W_\Even \wedge V_\Even \wedge S$ \tcp*{restrict strategy to winning region}
		$S_\Odd \gets W_\Odd \wedge V_\Odd \wedge S$ \tcp*{restrict strategy to winning region}
        \Return $W_\Even, W_\Odd, S_\Even, S_\Odd$ \;
	}
	\BlankLine
	\Def{{\textup{$\texttt{onestep}_0$}}(X, Z)}{
		$Z' \gets \texttt{subst}(\texttt{even}(Z), \vec{x}\to\vec{x}')$ \tcp*{rename variables of even region to $\vec{x}'$}
		$A \gets V_\Even \land X \land \texttt{exists}(E \land Z', \vec{x}')$ \tcp*{even-to-even}
		$B \gets V_\Odd \land X \land \texttt{forall}(V_\Odd \land (\neg E \lor Z'), \vec{x}')$ \tcp*{odd-to-even, encoding of $u\to v \implies v\in Z \equiv \neg E \lor Z'$}
	    \Return $A \lor B$
	}
	\BlankLine
	\Def{\texttt{\textup{even}}(Z)}{
	    \Return $({V}_0 \wedge \neg Z) \vee ({V}_1 \wedge Z)$ \tcp*{even-priority and not in Z plus odd-priority and in Z}
	}
	\BlankLine
	\Def{\texttt{\textup{odd}}(Z)}{
	    \Return $({V}_0 \wedge Z) \vee ({V}_1 \wedge \neg Z)$ \tcp*{even-priority and in Z plus odd-priority and not in Z}
	}
    \caption{The DFI algorithm using BDD operations}
	\label{algo:dfi_bdd}
\end{algorithm}

Algorithm \ref{algo:dfi_bdd} outlines the symbolic implementation of \texttt{DFI}. The symbolic \texttt{DFI} algorithm uses $d+1$ BDDs plus $3$ additional BDDs to store essential information plus $2$ helper sets for subresults:
\begin{itemize}[nosep]
	\item the current estimation of distractions $Z$
	\item a $F_p$ set for every priority $p$ with $0\leq p\leq d$; so in total these are $d+1$ BDDs
	\item a set $S$ for the chosen strategies.
	\item sets $X$ and $W$ that store subresults.
\end{itemize}

Some procedures have been redesigned to better suit a symbolic implementation.
Lines 8-14 are the symbolic or set-based version of an explicit iteration of all vertices of priority $p$ that are not frozen.
There are different possiblities for computing the various sets.
For example, we could choose a different order of operations at line~8.
In our initial experiments, the presented algorithm performed best.
Similarly, we implement a $\texttt{onestep}_0$ method that only computes the vertices in $X$ (unfrozen and of priority $p$) that reach the region $X$ that is currently good for player Even.

To update the strategy for the unfrozen vertices of priority $p$, the explicit algorithm simply overwrites the current strategy in an \texttt{int} array with a single chosen successor.
Our symbolic approach is different, as the strategy is now a set of edges.
First we remove the current strategy (line~12).
Then we add all strategies for all vertices controlled by Even that can go to the current even region.
Finally we add all strategies for all vertices controlled by Odd that can go to the current odd region.
Notice now that the set $S$ can have multiple successors for a single vertex.
While we currently do not use this, in a fully symbolic LTL toolchain this may be quite beneficial, as it allows the synthesis algorithm to pick different valid strategies depending on which combination yields the most optimal controller.
This is a serious advantage that explicit parity game solving algorithms do not have.

There are a few more explicit iterations over vertices in the original algorithm of~\cite{DBLP:journals/corr/abs-1909-07659} that we can replace with set-based operations.
To freeze and thaw vertices, the original algorithm uses a simple iteration over vertices.
This is a bit more work using BDDs and again we have more options to play with the order in which operations are carried out.
See lines~16--21 in Algorithm~\ref{algo:dfi_bdd}.
There can be many different ways to compute the set of unfrozen vertices of a priority less than or equal to $p$.
We found that the presented method worked well, but there is room for improvement here, including precomputing the set $V^{<p}$, and using e.g. heuristics to determine the best order of operations at runtime.

%
%

Notice that we can remove lines 12--14 and 28--29 without affecting the rest of the algorithm.
In some cases, for example if we just want to solve realisability problems instead of full synthesis, it is not required to obtain a winning strategy.
The \texttt{DFI} algorithm can be run without computing the strategy.
Thus we have a variant of symbolic \texttt{DFI} with no strategy computation.

\subsection{Symbolic \texttt{FPJ}}

\begin{algorithm}[tbp]
    \Def{\texttt{\textup{fpj}}(PG)}{
		$Z \gets V_0$ \tcp*{start with ``Even wins all even-priority vertices''}
		$J \gets \emptyset$ \tcp*{start with empty justification graph}
		$U \gets V \land \neg \texttt{exists}(J, \vec{x}')$ \tcp*{compute unjustified vertices}
		\While(\tcp*[f]{until all vertices justified}){$U \neq \emptyset$}{
		    $Z, J \gets \texttt{next}(Z, J, U)$ \tcp*{update $Z$ and $J$}
			$U \gets V \land \neg \texttt{exists}(J, \vec{x}')$ \tcp*{recompute unjustified vertices}
		}
		
		$W_\Even \gets Z$ \tcp*{player Even wins vertices in $Z$}
		$W_\Odd \gets V \wedge \neg Z$ \tcp*{player Odd wins vertices not in $Z$}
		$S_\Even \gets J \wedge V_\Even \wedge W_\Even $ \tcp*{restrict strategy to winning region}
		$S_\Odd \gets J \wedge V_\Odd \wedge W_\Odd $ \tcp*{restrict strategy to winning region}
		\Return{$W_\Even, W_\Odd, S_\Even, S_\Odd$}
	}
	\BlankLine
	\Def{\texttt{\textup{next}}(Z, J, U)}{
		$p \gets 0$ \;
		\lWhile(\tcp*[f]{find lowest priority}){$V^p \land U = \emptyset$}{$p \gets p+1$}
	    $U \gets V^p \land U$ \tcp*{restrict $U$ to vertices of priority $p$}
	    $C \gets U \land \texttt{xor}(Z, \texttt{phi}(Z))$ \tcp*{compute unjustified $p$-vertices whose winning region changes}
	    
	    \If(\tcp*[f]{did anything Change?}){$C \neq \emptyset$}{
	        $R \gets \texttt{reaches}(J, C)$ \tcp*{compute justified vertices depending on vertices in $C$}
	        \If(\tcp*[f]{if $p$ is even...}){$p \bmod 2 = 0$}{
	            $Z_R \gets (Z \wedge \neg (R \land V_1)) \land (((V^0 \lor V^1)\dotsc)\lor V^{p-1})$ \tcp*{remove odd-priority vertices $<p$}
	        }
	        \Else(\tcp*[f]{otherwise...}){
	            $Z_R \gets (Z \vee (R \land V_0)) \land (((V^0 \lor V^1)\dotsc)\lor V^{p-1})$ \tcp*{reset even-priority vertices $<p$}
	        }
        	$V^{>p} \gets (((V^{p+1} \lor V^{p+2})\dotsc)\lor V^{d})$ \tcp*{compute $V^{>p}$}
	        $Z'\gets (Z \wedge V^{>p}) \vee \texttt{xor}(Z \land V^p, C) \lor Z_R$ \tcp*{update Even winning region $Z$}
	        $J' \gets (J \wedge \neg R) \vee \texttt{strategy}_\Diamond (Z', C)$ \tcp*{update justification graph (prune, then add)}
	    }
	    \Else(\tcp*[f]{did nothing Change?}){
	        $Z' \gets Z$ \tcp*{no change to $Z$}
	        $J'\gets J \vee \texttt{strategy}_\Diamond (Z', U)$ \tcp*{update justification graph (no prune, just add)}
	    }
	    
	    \Return{$Z', J'$}
	}
	\BlankLine
	\Def{\texttt{\textup{reaches}}(J, X)}{
	    $X' \gets \emptyset$ \tcp*{start with empty set}
	    \While(\tcp*[f]{until fixpoint...}){$X' \neq X$}{
 	        $X' \gets X$ \tcp*{update previous X}
	        $X \gets X \vee (\texttt{exists}(J \land \texttt{subst}(X, \vec{x}\to\vec{x}'), \vec{x}'))$ \tcp*{add preimage in $J$ of $X$ to $X$}
	    }
	    \Return $X$ \tcp*{return all vertices reaching $X$ via $J$}
	}
	\BlankLine
	\Def{\texttt{\textup{phi}}(Z)}{
	    \Return{$(X_\Even \wedge \Diamond Z) \vee (X_\Odd \wedge \Box Z)$} \tcp*{compute $\texttt{onestep}_0$ basically}
	}
	\BlankLine
	\Def{\textup{$\texttt{strategy}_\Diamond$}(Z, U)}{
	    $W_E \gets U \wedge {V}_\Diamond \wedge Z $ \tcp*{all Even-controlled vertices in $U$ that are won by Even}
	    $W_O \gets U \wedge {V}_\Box \wedge \neg Z $ \tcp*{all Odd-controlled vertices in $U$ that are won by Odd}
	    $L \gets U \wedge  \neg (X_0 \vee X_1) \wedge V $ \tcp*{(losing) vertices in $U$ but not in $X_0$ or $X_1$}
	    $Z' \gets \texttt{subst}(Z, \vec{x}\to\vec{x}')$ \tcp*{rename Even winning region to $\vec{x}'$ variables}	    
	    \Return $(W_E \wedge Z' \wedge {E}) \lor \ (W_O \wedge \neg Z' \wedge {E}) \lor \ (L \wedge {E})$ \tcp*{good edges when winning, else all}
	}
    \caption{The \texttt{FPJ} algorithm using BDD operations}
    \label{algo:fpj_bdd}
\end{algorithm}

Algorithm \ref{algo:fpj_bdd} outlines the symbolic implementation of \texttt{FPJ}. The symbolic \texttt{FPJ} algorithm uses $2$ additional BDDs to store essential information plus $3$ helper sets for subresults:
\begin{itemize}[nosep]
	\item the current estimation of the Even winning set $Z$
	\item the current justification graph $J$ (which is a set of edges)
	\item a set of currently unjustified vertices $U$ derived from $J$
	\item the set of changed vertices $C$
	\item the set of pruned vertices $R$
\end{itemize}


On the procedural side of the symbolic FPJ implementation, there are only few significant modifications.
Compared to DFI, the symbolic implementation of FPJ is much more similar to its original explicit counterpart because the pseudo-code provided by Lapauw et al. \cite{DBLP:conf/vmcai/LapauwBD20} is based on set-operations.
Therefore no major changes were required.
Only the method $\texttt{strategy}_\Diamond$ was changed, as the explicit algorithm iterates over all vertices in $U$ in order to differentiate behavior depending on the winner of the vertex, and whether the winner also owns the vertex.
This corresponds to the symbolic implementation of lines 40-44 of algorithm \ref{algo:fpj_bdd}.
BDDs are first computed for all three cases (\textit{Even} controls and wins, \textit{Odd} controls and wins, \textit{Even} or \textit{Odd} controls and loses).
Then the algorithm proceeds to compute winning moves for vertices that are controlled and won by the same player, and compute all outgoing edges for vertices lost by the player which controls it.

Notice that contrary to symbolic \texttt{DFI}, computing the winning strategies is now integrated into the algorithm.
There is therefore no version of \texttt{FPJ} with no strategy computation.

\section{Implementation of the symbolic algorithms}

We implemented the symbolic algorithms using Python and the BDD packages provided by the Python library \texttt{dd}\footnote{\url{https://github.com/tulip-control/dd}}.
This library offers the CUDD backend using Cython bindings.
Furthermore, we used the \texttt{LineProfiler} package\footnote{\url{https://github.com/pyutils/line\_profiler}} to analyse the performance of the algorithms and find optimisations as discussed below.
We assume that a BDD package implements the methods \texttt{exists}, \texttt{forall}, binary operators \texttt{and}, \texttt{or} and \texttt{xor}, the method \texttt{subst} for variable renaming, and \texttt{neg} to compute complement sets in constant time using complement edges.
These are trivial requirements fulfilled by typical BDD implementations.

The implementations of our symbolic algorithms and the scripts and data of the empirical experiments are available online~\footnote{\url{https://github.com/olijzenga/bdd-parity-game-solver/tree/v1.1}}.

\paragraph{Optimisations}

A few optimisations were tested using \texttt{LineProfiler}. \texttt{LineProfiler} allows for profiling specific functions, and creates an overview of time spent on each line of Python code. This allows for easy analysis of small variations in code. 

The order of application of BDD operations can greatly influence the practical performance and even resolve bottlenecks in memory usage if the peak number of BDDs is sufficiently smaller.
At several points in algorithm \ref{algo:dfi_bdd}, we combine some BDD ($A$) with another BDD ($B$) which is a combination of many smaller BDDs ($B_0 \dotsc B_i$) (i.e. $V_{<p}$ and the set of all frozen vertices).
We do this by adding BDDs $B_0 \dotsc B_i$ to $A$ one by one.
This avoids combining two large BDDs at once by adding smaller BDDs to one large BDD instead.
Using \texttt{LineProfiler}, we found performance improvements of 11\% and 20\% for lines 8 and 17 of algorithm \ref{algo:dfi_bdd} for the \textit{full arbiter unreal 3} case from the empirical evaluation in section \ref{section:empirical}, which is the second largest BDD of the used benchmark sets.

In the symbolic \texttt{FPJ} algorithm, we need the set of unjustified vertices several times.
We compute this set once per iteration (lines~4,7 in Algorithm~\ref{algo:fpj_bdd}).
Furthermore, we found that the order of operations at lines~40--42 appears to be most optimal.


\section{Empirical Evaluation}
\label{section:empirical}

\begin{table}[p]
    \centering
    \begin{tabular}{ |p{2.2cm}||p{1.1cm}|p{1.1cm}|p{1.2cm}|p{1.5cm}|p{2.5cm}|p{2.5cm}| }
         \hline
          \textbf{Set} & \#games & avg. $n$ & max. $n$ & priorities & avg. out-degree & avg. \#bdd nodes  \\
         \hline
         \hline
         Lily               & 23  & 409   & 3047  & 3-8 & 1.47 & 1064   \\
         AMBA               & 8   & 2461  & 18635 & 3-4 & 1.43 & 1780  \\
         ltl2dba            & 58  & 921   & 31717 & 4-7 & 1.49 & 1876   \\
         Arbiters           & 15  & 2650  & 20928 & 4   & 1.65 & 6068  \\
         Detector           & 2   & 98    & 120   & 4   & 1.47 & 373     \\
         Buffer             & 2   & 14    & 17    & 4   & 1.30 & 72       \\
         Load balancer      & 11  & 1149  & 4712  & 3-8 & 1.65 & 2589   \\
         \hline
    \end{tabular}
    \caption{Statistics of benchmark sets used for empirical evaluation where $n$ is the number of vertices}
    \label{tbl:experiments}
\end{table}

\begin{table}[p]
    \centering
    \begin{tabular}{|p{2.6cm}||rr|rr||rr|rr|}
        \hline
        \textbf{Set}    & \multicolumn{4}{c||}{computes strategy} & \multicolumn{4}{c|}{no strategy} \\
        & \multicolumn{2}{c}{\texttt{dfi}} & \multicolumn{2}{c||}{\texttt{fpj}} & \multicolumn{2}{c}{\texttt{dfi-ns}} & \multicolumn{2}{c|}{\texttt{zlk}} \\
        \hline
        \hline
        Lily            & {0.38} & \stdevtext{8.1}    & 0.54  & \stdevtext{10.0}  & \textcolor{blue}{\textbf{0.19}}  & \stdevtext{6.3}    & 0.39     & \stdevtext{10.8}  \\
        AMBA            & {0.39} & \stdevtext{4.5}    & 0.70 & \stdevtext{4.0}   & \textcolor{blue}{\textbf{0.29}}  & \stdevtext{3.7}    & 0.63   & \stdevtext{9.4}   \\
        ltl2dba         & 21.11  & \stdevtext{11.4}   & {16.13} & \stdevtext{10.5}  & \textcolor{blue}{\textbf{14.63}} & \stdevtext{12.6}   & 15.08  & \stdevtext{12.0}  \\
        Arbiters        & 48.17   & \stdevtext{2.8}    & {23.98} & \stdevtext{2.7}   & 32.48                   & \stdevtext{5.0}    & \textcolor{blue}{\textbf{17.54}}  & \stdevtext{7.8}   \\
        Detector        & {0.0066} & \stdevtext{5.3}    & 0.0068                     & \stdevtext{4.3}   & \textcolor{blue}{\textbf{0.0046}}  & \stdevtext{3.6}    & 0.0047                     & \stdevtext{8.5}   \\
        Buffer          & 0.00082 & \stdevtext{4.2}    & {0.00058}  & \stdevtext{5.6}   & \textcolor{blue}{\textbf{0.00052}}  & \stdevtext{5.2}    & 0.00057   & \stdevtext{7.6}   \\
        Load\-balancer  & {0.50} & \stdevtext{2.5}    & 0.91  & \stdevtext{7.1}   & \textcolor{blue}{\textbf{0.37}}  & \stdevtext{13.4}   & 0.62    & \stdevtext{3.1}   \\
        \hline
    \end{tabular}
    \caption{Cumulative time in sec. average over 5 runs used to solve all games.}
    \label{tbl:results_time}
\end{table}

\begin{table}[p]
    \centering
    \begin{tabular}{|p{2.5cm}||r|r||r|r|}
        \hline
        \textbf{Set}    & \multicolumn{2}{c||}{computes strategy} & \multicolumn{2}{c|}{no strategy} \\
        & \multicolumn{1}{c}{\texttt{dfi}} & \multicolumn{1}{c||}{\texttt{fpj}} & \multicolumn{1}{c}{\texttt{dfi-ns}} & \multicolumn{1}{c|}{\texttt{zlk}} \\
        \hline
        \hline
        Lily            & {41647}  & 51747                   & \textcolor{blue}{\textbf{37431}}   & 44595                   \\
        AMBA            & {34299}  & 36261                   & \textcolor{blue}{\textbf{28143}}   & 34729                   \\
        ltl2dba         & {212242} & 263743                  & \textcolor{blue}{\textbf{183830}}  & 326327                  \\
        Arbiters        & {172235} & 231625                  & \textcolor{blue}{\textbf{153008}}  & 218382                  \\
        Detector        & 1400                      & {1393} & \textcolor{blue}{\textbf{1211}}    & 1427                    \\
        Buffer          & 244                       & 244                     & \textcolor{blue}{\textbf{222}}     & 253                     \\
        Load balancer  & {50411}  & 66389                   & \textcolor{blue}{\textbf{44407}}   & 60264                   \\
        \hline
    \end{tabular}
    \caption{Average Peak Live BDD Nodes (peak nodes per algorithm run)}
    \label{tbl:results_mem}
\end{table}

The empirical evaluation aims to study the performance of the discussed symbolic \texttt{DFI} and \texttt{FPJ} algorithms.
We compare four different symbolic algorithms:
\begin{itemize}[nosep]
	\item \texttt{dfi}: the standard symbolic \texttt{DFI} algorithm
	\item \texttt{dfi-ns}: symbolic \texttt{DFI} with no strategy computation
	\item \texttt{fpj}: symbolic \texttt{FPJ}
	\item \texttt{zlk}: symbolic Zielonka's recursive algorithm, provided by Sanchez et al~\cite{willemse}.
	
\end{itemize}

The main goals of the evaluations are to see whether the algorithms that compute strategies are competitive with state-of-the-art work, and furthermore to see whether not computing winning strategies influences the performance of symbolic \texttt{DFI}.

%
%

We evaluate the symbolic algorithms using the benchmark sets from the PGAME realizability track of SYNTCOMP 2020.
More detailed information on these benchmarks can be found in \cite{DBLP:journals/corr/abs-1711-11439}.
The SYNTCOMP benchmarks are parity automatons in extended HOA format~\cite{DBLP:journals/corr/abs-1912-05793} which were converted to explicit parity games in \texttt{PGSolver} format~\cite{pgsolver} by Knor\footnote{\url{https://github.com/trolando/knor}}, which is one of the participant tools in the 2020 edition of SYNTCOMP.
Contrary to some of the studies in the literature, we avoid using random games as we do not think that random games are representative for practical parity games.

Metrics of the used benchmark sets are shown in Table~\ref{tbl:experiments}. For the parity games in each set we have the average and highest number of vertices, priorities and the average outgoing degree. The average outgoing degree is the average of $|E|/|V|$ for each parity game. 
Comparing these two metrics gives an indication of the compression ratio of the BDDs. For example, the \textit{AMBA} benchmark set has a high compression ratio because only few BDD nodes are used to represent relatively many vertices and edges. 

Only benchmarks from formal verification and synthesis problems were used for the empirical evaluation.
Random parity games were not used, except to test the implementations.
In order to test our implementations, we used a combination of test games from the Oink~\cite{DBLP:conf/tacas/Dijk18} distribution as well as many randomly generated games.
The implementations of our symbolic algorithms and the scripts and data of the empirical experiments are available online.

In Table~\ref{tbl:results_time} and Table~\ref{tbl:results_mem}, we see the results for all benchmark sets. The provided times are cumulative for the entire benchmark set. The relative standard deviation (shown in parenthesis) is computed as a percentage of the mean time per parity game, and displayed as an average for each benchmark set. Testing was done on an Intel Core i7-7700HQ CPU @ 2.80GHz in conjunction with 16GB of RAM, running Python 3.6.9 on Ubuntu 18.04.
None of the algorithms used multi-threading.
The best results are highlighted in blue. 
The recorded times do not include the time for converting the original explicit parity game to a symbolic parity game, but only the difference between the four algorithms.

The first observation is that the implemented algorithms have a very similar performance compared to our reference symbolic solver, symbolic Zielonka.
There is only one benchmark class where Zielonka is the fastest solver, although this is simultaneously the benchmark class where all solvers require most time.
As one might expect, \texttt{dfi-ns} and \texttt{zlk}, which do not compute winning strategies, were almost always faster than the algorithms that do compute winning strategies.
We also notice that \texttt{dfi} and \texttt{dfi-ns} are typically faster than \texttt{fpj}, but not for the \textit{parameterized arbiter} specifications, where \texttt{fpj} outperformed \texttt{dfi} and \texttt{dfi-ns} for the \textit{full arbiter} cases, including one of 20928 vertices.
Further study might reveal why the \texttt{fpj} algorithm is particularly effective for these cases.

To give an idea of the memory usage of the evaluated algorithms, the peak number of live BDD nodes was recorded for each run, using CUDD's reporting of the peak number of nodes.
In table \ref{tbl:results_mem}, we see that \texttt{dfi-ns} consistently used the least amount of BDD nodes.
\texttt{dfi} used fewer BDD nodes than \texttt{zlk} for all benchmark sets.
Zielonka constructs subgames which are copies of the original games, which causes Zielonka's algorithm to be less memory efficient despite not computing winning strategies.
\texttt{fpj} also uses fewer BDD nodes than \texttt{zlk} for some cases, but this difference is neither as consistent nor as significant as it is between \texttt{dfi} and \texttt{zlk}.
We  conclude that a symbolic implementation of \texttt{dfi} may be a suitable choice when memory-usage must be as low as possible, although the differences are quite small.

To obtain insight into where \texttt{dfi}, \texttt{dfi-ns} and \texttt{fpj} spend most of their time, \texttt{LineProfiler} was again used.
This showed that all profiled algorithms were consistently using 90-99\% of their time on preimage computations. Both DFI and FPJ use these computations to update their estimate of the winning area (\texttt{onestep} and \texttt{phi} functions respectively), and to compute winning moves.
Preimage computations consist of several steps, all resulting in intermediate BDDs rather than doing all computations at once and then producing one final optimised and reordered result.

In general, the \texttt{dfi-ns} algorithm performs best across most of the evaluated benchmark sets. This shows that DFI can be a good alternative to Zielonka for both explicit (as shown in \cite{DBLP:journals/corr/abs-1909-07659}) and symbolic parity games. For the strategy derivation algorithms \texttt{dfi} and \texttt{fpj} there is no clear winner. Both \texttt{dfi} and \texttt{dfi-ns} have shown to be most memory efficient, especially when compared to Zielonka. In terms of time efficiency, it is hard to draw any major conclusions from our empirical evaluation as it remains hard to predict when an algorithm will out-perform the other, and different benchmark sets have entirely different properties and structures.

\section{Conclusion and discussion}
\label{section:conclusion}
We have proposed and implemented two symbolic parity game algorithms that yield winning strategies.
This is an important step towards fully symbolic LTL synthesis via parity games, which requires the winning strategies in order to construct controllers.
Our evaluation on real practical games derived from LTL synthesis demonstrates that the method is as fast as other symbolic parity game solvers, while producing winning strategies.

This paper represents a first step into symbolic algorithms that derive strategies and there is plenty of room to improve.
Most interesting would be to look at the encoding and variable reordering, but this is also the most difficult direction as we would need to consider the full LTL synthesis toolchain, including symbolic encoding of LTL to parity automata.
In this work, we used a simple variable ordering that places all pre-variables before all post-variables; alternatives such as interleaved variable orderings which are common in model checking would be recommended as a first step to try.
Easier improvements can be found by considering crafting specialised BDD operations that combine several operations in fewer steps.
In particular preimage computation dominates the runtime of our algorithms so any improvements to preimage computation immediately improves the entire algorithm.
Furthermore, we currently make use of sets $V^{>p}$ and $V^{<p}$ that could be precomputed.
We do not yet make use of parallel computation, as we used CUDD for this study, but if we switch to a BDD package like Sylvan, we could also find parallel speedups.
We did not investigate tuning CUDD with respect to automated variable reordering, which has shown to be beneficial for the related field of safety synthesis, although variable reordering tends to dominate runtimes of algorithms that depend on it.
We did not compare with explicit algorithms like the implementations in Oink, but our experiments (Table~\ref{tbl:results_time}) already show such small runtimes that a comparison is almost meaningless. For a proper comparison with explicit algorithms, we need examples of practical parity games that are much larger, actually challenging explicit parity game algorithms, and that might have some kind of structure that can be exploited by symbolic algorithms.
Current benchmark sets with large practical parity games unfortunately target explicit solvers and do not retain any structure of the original specifications.
Also, the implementations in Oink are highly optimized whereas our symbolic algorithms are not.
We could consider using multi-terminal BDDs to obtain additional compression, but the current BDDs are small enough that approaches using multi-terminal BDDs are currently not attractive.
Both \texttt{dfi} and \texttt{fpj} produce potentially many winning strategies, unlike explicit algorithms, since they do not select a single good successor when there are multiple good successors.
It may be interesting to compare the two algorithms with respect to the number of obtained strategies, if there is a significant difference.
Due to time and space constraints, we are unable to implement many of these ideas ourselves at this time.

\section*{Acknowledgements}
The second author of this paper is supported by the European Union's Horizon 2020 research and innovation programme under the Marie Sklodowska-Curie grant agreement No 893732.

\bibliographystyle{eptcs}
\bibliography{literature}

\clearpage
\newpage



\end{document}